\documentclass[twocolumn,aps,prb,nofootinbib,superscriptaddress,10pt]{revtex4-2}

\usepackage{amsmath}
\usepackage{amssymb}
\usepackage{graphicx}
\usepackage{hyperref}
\usepackage{newtxtext}
\usepackage[varvw]{newtxmath}
\usepackage{bbm}

\hypersetup{
    colorlinks,
    linkcolor={blue},
    citecolor={blue},
    urlcolor={blue}
}

\graphicspath{{./}{./figs/}}

\newcommand{\rme}{\mathrm{e}}
\newcommand{\rmi}{\mathrm{i}}
\newcommand{\rmd}{\mathrm{d}}
\newcommand{\const}{\mathrm{const.}}

\newcommand{\SO}{\mathrm{SO}}
\newcommand{\SU}{\mathrm{SU}}
\DeclareMathOperator{\Tr}{Tr}
\DeclareMathAlphabet{\mymathbb}{U}{BOONDOX-ds}{m}{n}

\begin{document}

\title{Phase diagrams of SO(\(\boldsymbol{N}\)) Majorana-Hubbard models:\\ Dimerization, internal symmetry breaking, and fluctuation-induced first-order transitions}

\author{Lukas Janssen}

\affiliation{Institut f\"ur Theoretische Physik and W\"urzburg-Dresden Cluster of Excellence ct.qmat, TU Dresden, 01062 Dresden, Germany}

\author{Urban F.\ P.\ Seifert}

\affiliation{Universit\'{e} de Lyon, ENS de Lyon, Universit\'{e} Claude Bernard, CNRS, Laboratoire de Physique, 69342 Lyon, France}
\affiliation{Kavli Institute for Theoretical Physics, University of California, Santa Barbara, CA 93106-4030, USA}

\begin{abstract}
We study the zero-temperature phase diagrams of Majorana-Hubbard models with SO($N$) symmetry on two-dimensional honeycomb and $\pi$-flux square lattices, using mean-field and renormalization group approaches.
The models can be understood as real counterparts of the SU($N$) Hubbard-Heisenberg models, and may be realized in Abrikosov vortex phases of topological superconductors, or in fractionalized phases of strongly-frustrated spin-orbital magnets.
In the weakly-interacting limit, the models feature stable and fully symmetric Majorana semimetal phases.
Increasing the interaction strength beyond a finite threshold for large $N$, we find a direct transition towards dimerized phases, which can be understood as staggered valence bond solid orders, in which part of the lattice symmetry is spontaneously broken and the Majorana fermions acquire a mass gap.
For small to intermediate $N$, on the other hand, phases with spontaneously broken SO($N$) symmetry, which can be understood as generalized N\'eel antiferromagnets, may be stabilized. These antiferromagnetic phases feature fully gapped fermion spectra for even $N$, but gapless Majorana modes for odd $N$.
While the transitions between Majorana semimetal and dimerized phases are strongly first order, the transitions between Majorana semimetal and antiferromagnetic phases are continuous for small $N \leq 3$ and weakly first order for intermediate $N \geq 4$.
The weakly-first-order nature of the latter transitions arises from fixed-point annihilation in the corresponding effective field theory, which contains a real symmetric tensorial order parameter coupled to the gapless Majorana degrees of freedom, realizing interesting examples of fluctuation-induced first-order transitions.
\end{abstract}

\date{November 2, 2021}

\maketitle

\section{Introduction}

Majorana fermions are hypothetical particles that constitute their own antiparticles. They represent real fermion modes that comprise half the degrees of freedom of the usual complex fermions~\cite{elliott15}.
In condensed matter systems, Majorana fermions can arise as effective excitations in quantum many-body systems~\cite{alicea12}. An illustrative example is given by a one-dimensional chain of spinless fermions in the presence of $p$-wave superconducting order, which features protected Majorana zero modes at its open ends~\cite{kitaev01}. In a similar way, Abrikosov vortices of a two-dimensional topological superconductor can host Majorana zero modes~\cite{fu08}, providing a way to realize two-dimensional systems of interacting Majorana fermions~\cite{rahmani19}. Importantly, the width of the Majorana bands can be tuned by a gate voltage in these systems, allowing one to access the regime of strong interactions between Majorana fermions even with weak underlying electron-electron interactions~\cite{chiu15}.
Besides applications in topological quantum computing~\cite{dassarma15}, two-dimensional Majorana fermion systems have therefore received interest as tunable platforms to investigate effects of strong interactions, spontaneous symmetry breaking, and quantum phase transitions~\cite{affleck17, wamer18, li18, li19, tummuru21}.

Majorana fermions can also emerge in insulating magnets in situations when magnetic frustration is significant. A well-known example is given by the Kitaev honeycomb model, which describes spins-$1/2$ localized on the sites of a honeycomb lattice and subject to bond-dependent nearest-neighbor exchange interactions~\cite{kitaev06}.
The model is exactly solvable using a parton decomposition, in which the spin Hamiltonian is mapped to a tight-binding Hamiltonian of Majorana fermions hopping in the background of a static $\mathbb Z_2$ gauge field. 
This construction has recently been extended to other tricoordinated lattices~\cite{yao07, yang07, mandal09, hermanns14, kimchi14, obrien16, krueger20, eschmann20}, as well as to systems with larger local Hilbert spaces~\cite{yao09, nakai12, natori20, chulliparambil20}.
In the latter cases, instead of a single Majorana fermion, an $N$-component vector of Majorana fermions emerges at each lattice site. 
This vector transforms in the fundamental representation of SO($N$), such that its components can be thought of as \textit{colors} of a global SO($N$) symmetry.

In this work, we study the phase diagrams of Majorana fermion models with generalized Heisenberg exchange interaction featuring $\SO(N)$ symmetry. These models fall into the larger class of Majorana-Hubbard models~\cite{rahmani19} and can be thought of as Majorana versions of the well-studied SU($N$) Hubbard-Heisenberg models that involve complex fermions~\cite{affleck88, marston89, read89, read90, assaad05, lang13}.
We hence dub these models SO($N$) Majorana-Hubbard models. 
Apart from their intrinsic appeal of constituting a systematic family of models with generalized exchange interactions, some instances at small $N$ are directly applicable to the low-energy physics of frustrated quantum magnets in two dimensions:
In particular, Ref.~\cite{seifert20a} proposed a spin-orbital generalization of the Kitaev honeycomb model with an additional Heisenberg exchange interaction, yielding a theory of three Majorana fermions on the honeycomb lattice with $\SO(3)$-symmetric exchange interactions coupled to a static $\mathbb{Z}_2$ gauge field.
Fixing a gauge in the ground-state flux sector, one arrives at the $\SO(3)$ Majorana-Hubbard model discussed in this work.
The $\SO(3)$ Majorana-Hubbard model can alternatively also arise in a parton description of $S=1$ Heisenberg antiferromagnets.
Further, the $\SO(6)$ Majorana-Hubbard model could emerge as an effective theory in the large-$U$ limit of an $\SU(4)$-symmetric spin-orbital model with two particles per site, as the antisymmetric product of two fundamental representations of $\SU(4)$ is six-dimensional, see also Ref.~\cite{kesel20} for an explicit mapping.
In the large-$N$ limit, order-parameter fluctuations in the SO($N$) Majorana-Hubbard models are suppressed, similar to their complex SU($N$) counterparts~\cite{affleck88,marston89,read89,read90}, enabling a controlled mean-field analysis.

Here, we first use lattice mean-field theory to investigate the ground-state phase diagrams as function of interaction strength for different values of $N$.
We demonstrate the existence of three symmetry-distinct states in the phase diagrams of the models on the honeycomb and $\pi$-flux square lattices:
(1)~A symmetric Majorana semimetal phase at weak interactions for all values of $N$, featuring $N$ gapless Majorana modes at the nodal point in the half-Brillouin zone~\cite{seifert20a}.
(2)~A dimerized phase, in which part of the lattice symmetry is spontaneously broken. The order can be understood as a staggered valence bond solid~\cite{xu11}, and is stabilized for strong interactions and large values of $N$.
(3)~An $\SO(N)$-symmetry-broken phase, characterized by the symmetry breaking pattern $\SO(N) \to \SO(2) \otimes \dots \otimes \SO(2)$ with $N/2$ [$(N-1)/2$] factors of $\SO(2)$ for even (odd) values of $N$.
This state occurs at intermediate to strong interaction for small values of $N$, and can be understood as a generalized N\'eel antiferromagnet. For even $N$, all Majorana modes are gapped out, while for odd $N$, a single mode remains gapless.
In addition, for intermediate values of $N$, a fully gapped coexistence phase, characterized by both antiferromagnetic and dimerized order, can occur at strong interactions in the SO(3) honeycomb lattice model.

Secondly, we study the natures of the various transitions occurring in the phase diagrams.
In mean-field theory, the transitions towards the dimerized phase as function of interaction strength are strongly first order. As these occur at large $N$, and/or between ordered states that break different symmetries, we expect this conclusion to hold also upon the inclusion of quantum fluctuations beyond mean-field theory.
By contrast, the transitions between Majorana semimetal and antiferromagnetic phases, occurring for small values of $N$, are continuous on the mean-field level.
In order to study the effects of order-parameter fluctuations, which may become sizable for small $N$, we devise the corresponding low-energy effective field theories describing these transitions. These continuum field theories exhibit a unique upper critical spatial dimension of three, allowing us to reveal the universal properties of the transitions within a controlled $\epsilon$ expansion. 
We find that the semimetal-to-antiferromagnet transition remains continuous for $N\leq 3$, while it becomes weakly first order for $N \geq 4$ upon taking quantum fluctuations into account. The weakly-first-order nature of the transition for $N \geq 4$ can be understood to arise from a fixed-point-annihilation mechanism in the corresponding effective field theory, which contains a real symmetric tensorial order parameter coupled to the gapless Majorana degrees of freedom. This effect is similar (though reversed as function of $N$) to the situation in the Abelian Higgs model with $N$ complex boson fields in $4-\epsilon$ dimensions, which features a continuous transition on the mean-field level that is believed to become weakly first order for $N$ below a certain critical $N^\text{cr}(\epsilon) = 182.95(1 - 1.752\epsilon + 0.798 \epsilon^2 + 0.362 \epsilon^3) + \mathcal O(\epsilon^4)$ as a consequence of a fixed-point annihilation~\cite{halperin74, nahum15, ihrig19}. A similar fixed-point annihilation has recently been discussed in a number of similar relativistic~\cite{gies06, kaplan09, braun14, herbut16b, herbut16a, gracey18, ma19, nahum20, ma20} and nonrelativistic~\cite{herbut14, janssen17} field theories in different dimensions. It leads to an exponentially large, but finite, correlation length and a quasiuniversal regime characterized by approximate power laws in various observables.

The remainder of this work is organized as follows:
In the next section, we introduce the SO($N$) Majorana-Hubbard models and discuss its ground states in the limiting cases for weak and strong interactions, respectively.
The phase diagram as obtained from lattice mean-field theory is presented in Sec.~\ref{sec:MFT}.
Section~\ref{sec:RG} contains a discussion of the natures of the various quantum phase transition using a renormalization group analysis.
We conclude in Sec.~\ref{sec:conclusions}.
Technical details are deferred to two appendices.

\section{Models}
\label{sec:model}

Motivated by the frustrated spin-orbital models studied in Ref.~\cite{seifert20a}, we consider $N$ colors of Majorana fermions located on the sites of a bipartite lattice, transforming in the fundamental representation of $\SO(N)$.
The SO($N$) Majorana-Hubbard Hamiltonian is comprised of a hopping term with hopping parameters $t_{ij}$ and an $\SO(N)$-symmetric nearest-neighbor interaction term with coupling constant $J$, which can be thought of as a generalized Heisenberg interaction for Majorana fermions,
\begin{align} \label{eq:ham}
    \mathcal{H} = \sum_{\langle i j \rangle} \rmi t_{ij} c_i^\top c_j + J \sum_{a<b}\sum_{\langle i j \rangle} \left(\tfrac{1}{2} c_i^\top L^{ab} c_i \right) \left(\tfrac{1}{2} c_j^\top L^{ab} c_j \right),
\end{align}
where $\langle i j \rangle$ denote nearest-neighbor bonds between adjacent sites $i$ and $j$ of the lattice. In order to fix the sign of $t_{ij}$, we assume that $i \in A$ and $j \in B$, where $A$ and $B$ are the two sublattices.
The spinor $c_i \equiv (c_i^{1}, \dots c_i^{N})^\top$ consists of $N$ colors of Majorana fermions, satisfying canonical anticommutation relations, $\{c_i^\alpha,c_j^\beta\} = 2\delta_{ij} \delta^{\alpha \beta}$, and hermiticity, $c_i^{\alpha\dagger} = c_i^{\alpha}$, $\alpha,\beta = 1,\dots,N$.
The $N(N-1)$ matrices $L^{ab} \in \mathbb C^{N\times N}$ with $1 \leq a < b \leq N$ and entries $(L^{ab})_{\alpha \beta} = -\rmi (\delta^{\alpha}_a \delta^{\beta}_b - \delta^{\beta}_a \delta^{\alpha}_b)$ are the $\SO(N)$ generators in the fundamental representation. 

Using the above representation of $L^{ab}$ and evaluating the summation over $a$ and $b$ in Eq.~\eqref{eq:ham} explicitly, we can alternatively rewrite the interaction term as
\begin{equation} \label{eq:HJ_bonds}
    \mathcal{H}\Bigr|_{t=0} = \frac{J}{2} \sum_{\langle ij \rangle}  \left(c^\top_i c_j\right) \left(c^\top_i c_j\right) + \const,
\end{equation}
which can be viewed as a Majorana analog of the $\SU(N)$ extension of the Hubbard-Heisenberg interaction~\cite{affleck88,marston89}.
From Eq.~\eqref{eq:HJ_bonds}, it becomes clear that the Hamiltonian \eqref{eq:ham} enjoys an $\mathrm{O}(N)$ symmetry of rotations of the $N$-dimensional Majorana vector $c \mapsto O c$ for some orthogonal matrix $O$.
The symmetry group here also includes improper orthogonal transformations, which may possibly be no longer admissible once a local fermion parity constraint is introduced.%
\footnote{Note, however, that the itinerant Majoranas in Refs.~\cite{seifert20a} and \cite{chulliparambil21} are invariant under $\mathrm{O}(N)$ transformations as the local parity constraint in these models involves additional ``gauge'' Majoranas which can be used to absorb additional parity transformations. We refer to Ref.~\cite{chulliparambil21} for an extended discussion.}

\begin{figure}[tb]
    \includegraphics[width=\linewidth]{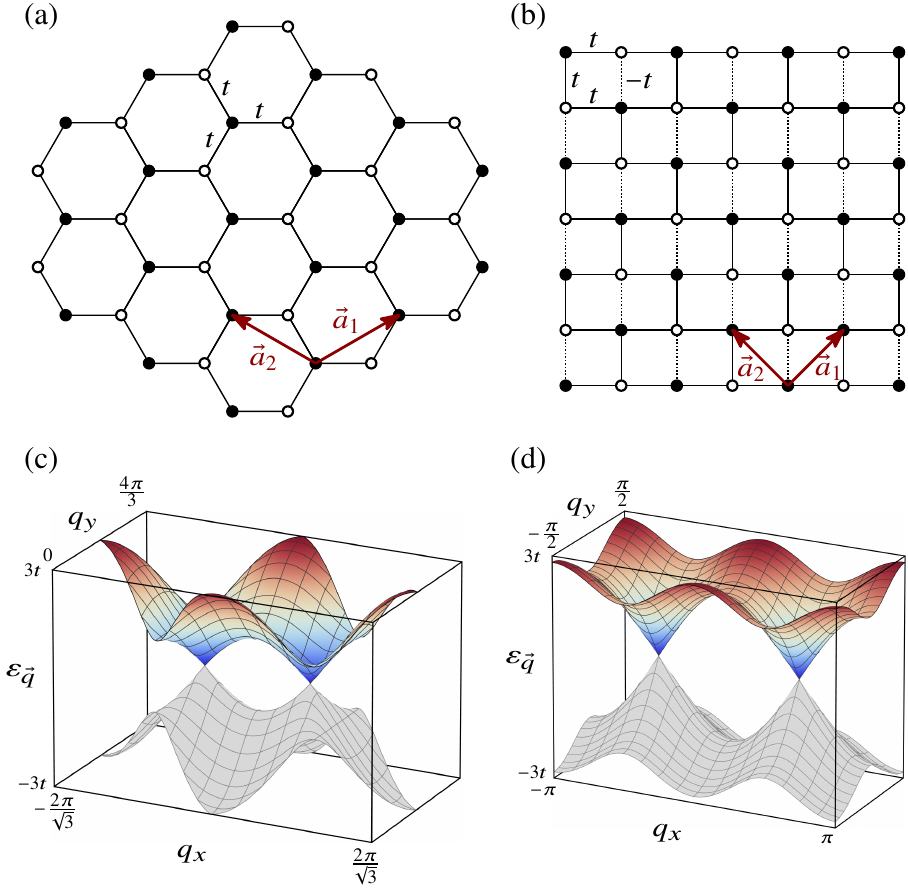}
    \caption{(a)~Honeycomb lattice with uniform hopping parameter $t_{ij} \equiv t$ and two sublattices $A$ and $B$, indicated by the black and white dots, respectively. 
    (b)~Same as (a), but for the $\pi$-flux square lattice with uniform (alternating) signs $t_{ij} \equiv t$ ($t_{ij} = \pm t$) along horizontal (vertical) bonds, as indicated by the solid (alternating solid and dashed) lines.
    (c)~Majorana dispersion for $J = 0$ in the honeycomb-lattice model with nodal points at $\vec q = \frac{2\pi}{3}(\pm\frac{1}{\sqrt{3}},1)$, using units in which the distance between nearest-neighbor sites is $a=1$.
    The shaded negative-energy band serves as a reminder for the fact that excitations in the ``particle'' and ``hole'' bands are not independent, as Majorana fermions constitute their own antiparticles.
    (d)~Same as (c), but for the $\pi$-flux model with nodal points at $\vec q = (\pm\frac{\pi}{2}, 0)$.}
    \label{fig:lattices}
\end{figure}

In this work, we consider the model \eqref{eq:ham} on the honeycomb and $\pi$-flux square lattices. On the honeycomb lattice, we assume uniform hopping $t_{ij} \equiv t$ on all bonds $\langle ij \rangle$, see Fig.~\ref{fig:lattices}(a). On the $\pi$-flux square lattice, by contrast, we assume that a flux of $\pi$ is penetrating each square plaquette.
This may be represented, for instance, by $t_{ij} = t$ along horizontal bonds and alternating signs $t_{ij} = \pm t$ along vertical bonds~\cite{otsuka16, affleck17}, see Fig.~\ref{fig:lattices}(b).
Note that in the context of lattice gauge theory, our particular implementation of $\pi$-flux pattern on the square lattice corresponds to only one of many gauge-equivalent configurations. While our above choice explicitly breaks some of the square lattice's symmetries, these symmetries are restored in the lattice gauge theory by noting that physical symmetry operations act projectively and are to be paired with appropriate gauge transformations for explicit invariance \cite{wen02}.
For small values of $N$, members of the above family of models can be mapped to known models discussed in previous works:
For $N=3$, the honeycomb-lattice model emerges in the ground-state flux sector of a Kitaev-Heisenberg-type spin-orbital model~\cite{seifert20a}, and maps to the spin-orbital liquid of Refs.~\cite{yao09, natori20} in the weakly-interacting limit.
Similarly, for $N=2$, the $\pi$-flux model describes the ground-state flux sector of a generalized Kitaev spin-orbital liquid on the square lattice~\cite{nakai12, chulliparambil20}, perturbed by an additional Ising spin-spin interaction~\cite{seifert20a}.
In these spin-orbital realizations of the $\SO(2)$ and $\SO(3)$ Majorana-Hubbard Hamiltonians, the hopping term corresponds to a generalized Kitaev spin-orbital exchange coupling~\cite{chulliparambil20}, while the interaction terms map to Heisenberg and Ising spin-spin interactions, respectively~\cite{seifert20a}.
In the $\SO(2)$ model, the two Majorana modes can be combined into a single complex fermion $f_i = (c_i^1 + \rmi c_i^2)/2$. In this representation, the $\SO(2)$ Majorana-Hubbard Hamiltonian describes a model of spinless complex fermions subject to nearest-neighbor repulsion~\cite{seifert20a}, which has been intensely studied previously on both the honeycomb and $\pi$-flux lattices~\cite{wang14, li15, motruk15, capponi15, scherer15, hesselmann16, huffman17, huffman20, schuler21}.
The mapping can be understood as a consequence of the fact that $\SO(2)$ has only a single generator $L^{12} = \sigma^y$, the antisymmetric $2\times 2$ Pauli matrix, leading to a density-density form of the interaction.
We furthermore note that the zero-flux and $\pi$-flux configurations of the hopping parameters on the honeycomb and square lattices, respectively, satisfy the Grosfeld-Stern rule~\cite{grosfeld06}, and as such naturally emerge also in potential realizations of the models using Abrikosov vortex lattices of topological superconductors~\cite{liu15}.

In both the honeycomb and $\pi$-flux cases, the single-particle Majorana bands feature isolated nodal points at zero energy and a linear dispersion in the vicinity of the nodal points, see Figs.~\ref{fig:lattices}(c) and (d). 
The nodal points are protected by the internal and lattice symmetries of the Hamiltonian. In the honeycomb-lattice model, this includes a $2\pi/3$ rotational symmetry around a lattice site. On the $\pi$-flux square lattice, a $\pi/4$ rotation around a lattice site has to be paired with an appropriate local $\mathbbm Z_2$ transformation of the fermions in order to yield a symmetry of the model.
In the noninteracting limit, the $\SO(N)$ Majorana-Hubbard models on both lattices therefore realize Majorana semimetal phases that are stable against the inclusion of small interactions.
This is similar to the situation of the conventional Hubbard model on honeycomb and $\pi$-flux lattices~\cite{sorella92, assaad13, otsuka16}. Note, however, that the number of fermion excitations here is halved in comparison with the corresponding complex-fermion models. ``Particle'' and ``hole'' bands are not independent and in order to avoid double counting, one has to restrict the mode summations to either only half of the energies, say $\varepsilon_{\vec q} \geq 0$, or only half of the Brillouin zone momenta, say $q_x \geq 0$~\cite{seifert20a}.

The fate of the systems at large couplings $J/t$ is less obvious:
On one hand, the form of the interaction term as given in Eq.~\eqref{eq:ham} suggests an $\SO(N)$-symmetry-broken ground state with a staggered ordered parameter $\phi_j^{ab} = \langle \frac{1}{2} c_i^\top L^{ab} c_i \rangle$ for $J/t\gg 1$. Such a state can be understood as an $\SO(N)$ generalization of the usual $\SO(3)$-breaking N\'eel antiferromagnet.
In fact, in the case of $N=3$, Eq.~\eqref{eq:ham} can be perceived as a Majorana parton representation of a spin-1 antiferromagnetic Heisenberg Hamiltonian, which also features a N\'eel-ordered ground state on a bipartite lattice.
On the other hand, the interaction \eqref{eq:HJ_bonds} is readily rewritten as a quadratic form in terms of dynamic bond variables $c_i^\top c_j$, suggesting a mean-field decoupling in terms of the static mean fields $\chi_{ij} \propto \rmi \langle c_i^\top c_j \rangle$.
Finite mean fields $\chi_{ij}$ in general indicate the hybridization of Majorana fermions on different sites.
A varying configuration of the $\chi_{ij}$ on otherwise symmetry-equivalent bonds indicates a ground state that breaks lattice rotational and/or translational symmetries. In particular, $\chi_{ij}$ being finite on just one of the nearest-neighbor bonds adjacent to a given lattice site is suggestive of the forming of a dimer that is a singlet under $\SO(N)$.
While previous infinite density matrix renormalization group calculations for the case of $N=3$ on the honeycomb lattice~\cite{seifert20a} indeed provide evidence for the presence of an $\SO(N)$-breaking N\'eel-type ordered state for $J/t$ above a certain finite threshold, the mean-field decoupling of Eq.~\eqref{eq:HJ_bonds}, leading to $\SO(N)$-symmetric phases, becomes exact in the limit $N \to \infty$ \cite{affleck88,marston89}.
We therefore expect the phase diagram as function of $N$ and interaction strength $J$ to feature at least three phases: 
(1)~A Majorana semimetal phase for small $J/t$, (2)~a N\'eel antiferromagnetic phase for small $N$ and $J/t$ above a certain finite threshold, and (3)~an $\SO(N)$ symmetric phase that breaks lattice symmetries only.
In the next section, we use mean-field theory to map out the zero-temperature phase diagram in detail.

\section{Phase diagrams}
\label{sec:MFT}

\begin{figure*}[tb]
\includegraphics[width=\textwidth]{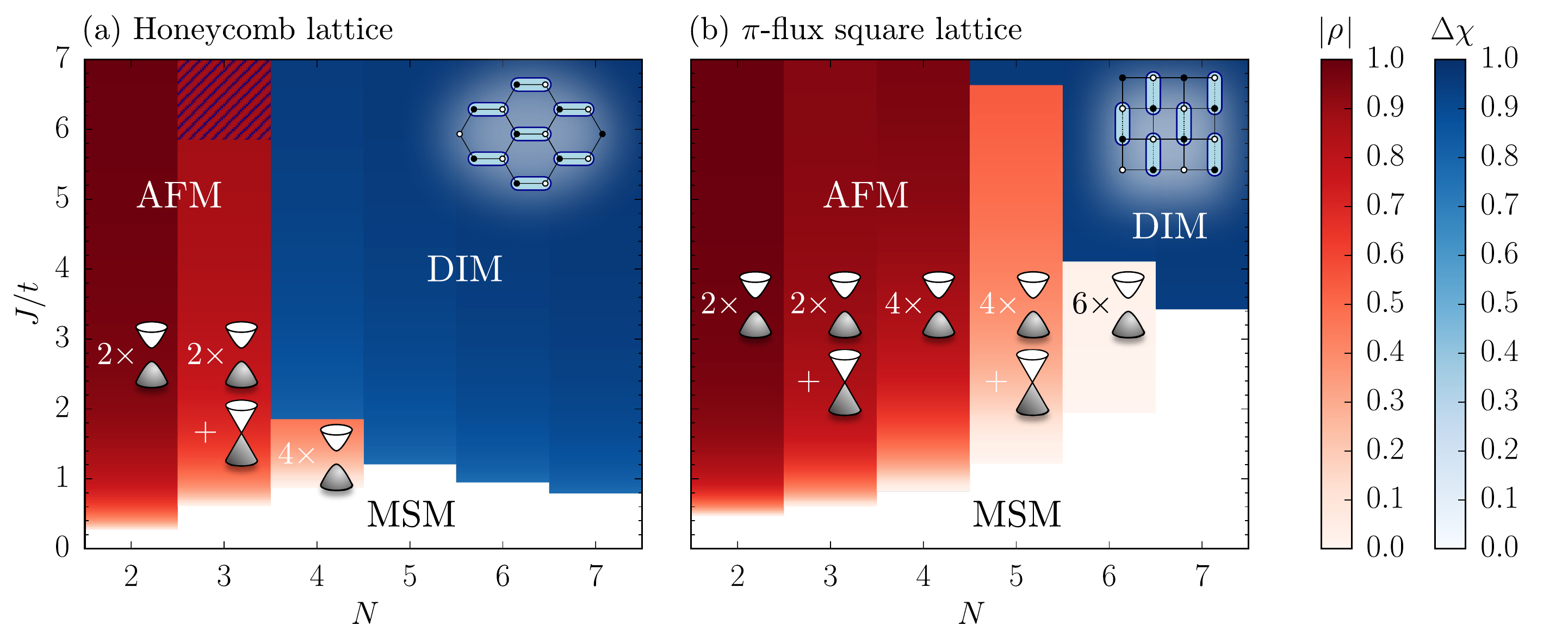}
\caption{Mean-field phase diagrams of $\SO(N)$ Majorana-Hubbard models as function of $J/t$ for different values of $N$ on (a)~honeycomb and (b)~$\pi$-flux square lattices. 
The red shading indicates the norm $|\rho|$ of the $\SO(N)$-breaking order parameter matrix $\rho = \sum_{a<b} \phi^{ab} L^{ab}$
, representing the generalized N\'eel antiferromagnetic order (AFM). This phase features a gapped (gapless) Majorana spectrum for even (odd) $N$, as illustrated in the insets.
The blue shading indicates an anisotropy in the bond order parameters, $\Delta \chi = \max \chi_{ij} - \min \chi_{ij}$, representing the gapped staggered dimerized phase (DIM). This phase can also be understood as a staggered valence bond solid, as illustrated in the inset.
The white region corresponds to the symmetric gapless Majorana semimetal phase (MSM).
The transitions involving the DIM phase are strongly first order, while the transition between MSM and AFM is continuous on the mean-field level. Renormalization group arguments presented in Sec.~\ref{sec:RG} indicate that the MSM-AFM transition remains continuous for $N \leq 3$, but becomes weakly first order for $N \geq 4$ upon the inclusion of quantum fluctuations.
For $N=3$ and large coupling $J/t \gtrsim 5.88$ in the honeycomb-lattice model, mean-field theory suggest a coexistence phase characterized by both antiferromagnetic and dimerized order [hatched region in (a)].}
\label{fig:pd}
\end{figure*}

\subsection{Mean-field ansatz}

To capture the competition between disordered, bond-ordered, and $\SO(N)$-symmetry-broken states, we perform an unrestricted Hartree-Fock decoupling of the interaction term as
\begin{multline} \label{eq:decoupling}
\sum_{a<b}\left(\tfrac{1}{2} c_i^\top L^{ab} c_i \right) \left(\tfrac{1}{2} c_j^\top L^{ab} c_j \right) 
\mapsto \\
\sum_{a<b}\left[\phi^{ab}_i \left(\tfrac{1}{2} c_i^\top L^{ab} c_i\right)
+ \left(\tfrac{1}{2} c_j^\top L^{ab} c_j\right) \phi^{ab}_j 
- \phi^{ab}_i \phi^{ab}_j \right]
\\
-  \rmi (N-1) \chi_{ij} c_i^\top c_j + \frac{N(N-1)}{2} \chi_{ij}^2,
\end{multline}
where $\phi^{ab}_i = \langle \frac{1}{2} c_i^\top L^{ab} c_i \rangle$ are $\SO(N)$-symmetry-breaking mean fields, and $\chi_{ij} = \langle \rmi c_i^\top c_j \rangle / N$ correspond to the bond variables.
One virtue of this decoupling is that in the $\SO(N)$-symmetric channel it reproduces to leading order in $1/N$ the Hubbard-Stratonovich decoupling of Eq.~\eqref{eq:HJ_bonds} akin to the $\SU(N)$ case~\cite{affleck88, marston89}.
As is well known, in the limit $N\to\infty$, the Hubbard-Stratonovich decoupling and the subsequent saddle-point approximation becomes exact \cite{affleck88,marston89}.

In the following, we assume that the translational symmetries remain unbroken, so that the $\SO(N)$-symmetry-breaking mean fields $\phi^{ab}_i$ depend only on the sublattice index $i\in A$ or $i \in B$.
We emphasize that this restriction excludes states with enlarged unit cells.
We allow for a possible (projective) rotational symmetry breaking, such that the bond mean-fields $\chi_{ij} \in \{\chi_1, \chi_2,\chi_3\}$ ($\chi_{ij} \in \{\chi_1, \chi_2,\chi_3,\chi_4\}$) may take different values on the three (four) distinct bonds in the unit cell of the honeycomb ($\pi$-flux) lattice.
In total, the problems on the honeycomb and $\pi$-flux lattices thus involve $N(N-1) + 3$ and $N(N-1) + 4$, respectively, mean fields, which we determine self-consistently.
To this end, we introduce a Fourier representation of the Majorana modes as $c^\alpha_{s,i} = N^{-1}_\mathrm{u.c.} \sum_{\vec k \in \mathrm{BZ}/2} [c^\alpha_{s}(\vec k) \rme^{\rmi \vec k \cdot \vec x} + c_{s}^{\alpha\dagger}(\vec k) \rme^{-\rmi \vec k \cdot \vec x}]$ with $s=A,B$ denoting the sublattice index, $N_\text{u.c.}$ the number of unit cells, and the $\vec k$-space summation extending over half of the Brillouin zone ($\mathrm{BZ}/2$). 
The mean-field Hamiltonian in momentum space is readily diagonalized on a finite-size lattice. We employ a momentum-space discretization of up to $36 \times 36$ unit cells.
We furthermore employ a small, but finite, temperature $T=0.08t$ to ensure numerical stability, and we have verified that our results are converged upon varying temperature and system size.
The self-consistent solution can then be found iteratively.
Note that due to the Majorana nature of the fermionic modes, the mean-field Hamiltonian is explicitly particle-hole symmetric on $\mathrm{BZ}/2$, such that no Lagrange multipliers are required to enforce a particular filling.

\subsection{Mean-field ground states}

The resulting mean-field phase diagrams for the $\SO(N)$ Majorana-Hubbard models on the honeycomb and $\pi$-flux lattices are shown in Fig.~\ref{fig:pd}. 
In it, the magnitudes of the mean-field parameters $|\rho| = |\sum_{a<b} \phi^{ab} L^{ab}|$ and $\Delta \chi = \max \chi_{ij} - \min \chi_{ij}$ as function of $J/t$ are depicted as blue and red color codings, respectively. Explicit forms of the mean-field parameters are shown in Appendix~\ref{app:mft}.
Here, we discuss the properties of the different phases.

\paragraph*{Majorana semimetal.}
In the weakly-interacting limit, the Majorana semimetal is stable for all values of $N$. 
Any instability occurs only for interactions beyond a finite threshold, which can be understood as a consequence of the linearly vanishing density of states at the Fermi level. 
The solution of the self-consistency equations is fully symmetric, featuring isotropic bond variables $(\chi_1, \chi_2, \chi_3) = (\chi,\chi,\chi)$ on the honeycomb lattice, and following the modulation of the hopping strength $(t,t,t,-t)$ on the $\pi$-flux lattice, i.e., $(\chi_1, \chi_2, \chi_3, \chi_4) = (\chi ,\chi ,\chi ,-\chi )$. Independent of $N$ and $J/t$ within this phase, we find $\chi = -0.52483184$ ($\chi = -0.47902212$) on the honeycomb ($\pi$-flux) lattice.
The low-energy spectrum in this phase is characterized by $N$ gapless Majorana modes at a unique nodal point in $\mathrm{BZ}/2$, as in the noninteracting cases displayed in Figs.~\ref{fig:lattices}(c) and (d).

\paragraph*{N\'eel antiferromagnetic order.}
For small $N \leq 4$ ($N \leq 6$) on the honeycomb ($\pi$-flux) lattice and increasing interaction strength, we find a direct phase transition from the Majorana semimetal to a staggered $\SO(N)$-symmetry breaking phase, which can be understood as the $\SO(N)$ generalization of the usual N\'eel antiferromagnet. 
On the level of mean-field theory, this transition is, independently of the value of $N$, continuous.
The presence of the generalized N\'eel antiferromagnet at intermediate to strong coupling and small $N$ is expected from the previous calculations for $N=3$ on the honeycomb lattice~\cite{seifert20a} and $N=2$ on both lattices~\cite{wang14, li15, motruk15, capponi15, scherer15, hesselmann16, huffman17, huffman20, schuler21}.
For general $N$, the spontaneous symmetry breaking of this antiferromagnetic state is ``maximal'' in the sense that the eigenvalues of $\phi^{ab} L^{ab}$ come in $\lfloor N/2 \rfloor$ identical pairs $\pm \tilde{\phi}$, where $\lfloor\,\cdot\,\rfloor$ corresponds to the floor function. For odd $N$, there is an additional zero  eigenvalue. 
The corresponding symmetry breaking pattern is
\begin{equation} \label{eq:pattern}
\mathrm{O}(N) \to
\begin{cases}
    \bigotimes^{N/2} \SO(2), & \text{for $N$ even}, \\
    \left[\bigotimes^{(N-1)/2} \SO(2) \right] \otimes \mathbb{Z}_2, & \text{for $N$ odd},
\end{cases}
\end{equation}
where the residual $\mathbb{Z}_2$ symmetry for $N$ odd corresponds to a $\pi$ phase rotation of the Majorana zero mode in the antiferromagnetic state.
The unbroken $\SO(2)$ subgroups become evident by noting that the Cartan generators $H^p$ of $\SO(N)$ with $p=1,\dots,\lfloor N/2 \rfloor$ lie in the broken-symmetry manifold~\cite{chulliparambil21}, which are block diagonal with
\begin{align} \label{eq:cartan-a}
H^p = 
\underbrace{\mymathbb{0}_2 \oplus \dots \oplus \mymathbb{0}_2}_{\text{$p-1$ blocks}} \mathrel{\oplus} \sigma^y \oplus 
\underbrace{\mymathbb{0}_2 \oplus \dots \otimes \mymathbb{0}_2}_{\text{$\frac{N}{2} -p$ blocks}}\,,
\end{align}
for $N$ even, and
\begin{align} \label{eq:cartan-b}
H^p = 
\underbrace{\mymathbb{0}_2 \oplus \dots \oplus \mymathbb{0}_2}_{\text{$p-1$ blocks}} \mathrel{\oplus} \sigma^y \oplus 
\underbrace{\mymathbb{0}_2 \oplus \dots \otimes \mymathbb{0}_2}_{\text{$\frac{N-1}{2} -p$ blocks}} \mathrel{\oplus} \mymathbb{0}_1\,, 
\end{align}
for $N$ odd.
Here, $\mymathbb{0}_1$ and $\mymathbb{0}_2$ denote one- and two-dimensional zero blocks, respectively, and $\sigma^y$ is the imaginary antisymmetric Pauli matrix.
For any $\SO(N)$-symmetry-breaking configuration $\phi^{ab}$, we can use a $O \in \SO(N)$ transformation to write
\begin{align}
    O^\top \phi^{ab} L^{ab} O = \tilde{\phi}^p H^p,
\end{align}
where above notion of ``maximal'' breaking leads to identical $\tilde{\phi}^p \equiv \tilde{\phi}$.
Considering the mean-field Hamiltonian following from Eq.~\eqref{eq:decoupling} in a $2N$-component-spinor notation $\Psi = \left( c^{1}_{A}, c^{1}_{B},c^{2}_{A}, c^{2}_{B}, \dots, c^{N}_{A}, c^{N}_{B}\right)^\top$, a staggered order in $\phi^{ab}$ is seen to generate a mass term for the Majorana fermions of the form $\Psi^\dagger \left(\phi^{ab} L^{ab} \otimes \tau^z \right) \Psi$, where the $2\times 2$ diagonal Pauli matrix $\tau^z$ acts on the sublattice degree of freedom.
With above transformation to the Cartan basis it becomes clear that there is a $\SO(2)$ freedom of mixing the two Majorana modes in each $2 \times 2$ block.
These considerations also show that for even $N$, the $\SO(N)$-broken antiferromagnetic state is fully gapped, while there remains a single gapless mode for odd $N$.

\paragraph*{Staggered dimerized order.}
On the other hand, for large $N \geq 5$ ($N \geq 7$) on the honeycomb ($\pi$-flux) lattice, we find a strong first-order transition from the semimetallic phase to a bond-ordered dimerized state that is fully $\SO(N)$ symmetric.
Instead, the bond mean fields $\chi_{ij}$ in the latter state break part of the lattice symmetry:
On the honeycomb ($\pi$-flux) lattice, we find one ``strong'' and two (three) ``weak'' bonds per unit cell, with $(\chi_1,\chi_2,\chi_3) = (\tilde\chi - \Delta\chi, \tilde\chi, \tilde\chi)$ [$(\chi_1,\chi_2,\chi_3,\chi_4) = (\tilde\chi - \Delta\chi, \tilde\chi, \tilde\chi, -\tilde\chi)$ or $(\tilde\chi , \tilde\chi, \tilde\chi, -\tilde\chi+ \Delta \chi)$], or symmetry-equivalent, with $0 < \Delta\chi < 1$ and some renormalized value $\tilde\chi<0$.
This configuration is adiabatically connected to the ``fully dimerized'' configuration, for which $(\chi_1,\chi_2,\chi_3) = (-1,0,0)$ [$(\chi_1,\chi_2,\chi_3,\chi_4) = (-1,0,0,0)$], corresponding to $\Delta\chi \to 1^-$ and $\tilde\chi\to 0^-$, and can be understood as a staggered valence bond solid~\cite{xu11}.
Note that the staggered dimerized order does not enlarge the two-site unit cell, but instead breaks the $2\pi/3$ rotational symmetry (combined symmetry from $\pi/4$ rotation and local $\mathbbm Z_2$ transformation) on the honeycomb ($\pi$-flux) lattice.
The lattice symmetry breaking fully gaps out the Majorana spectrum, rendering this phase insulating.

\paragraph*{Coexistent antiferromagnetic-dimerized order.}
Interestingly, the mean-field analysis of the $N=3$ model on the honeycomb lattice suggests for $J/t \gtrsim 5.88$ coexistent antiferromagnetic and dimerized order, characterized by both $\SO(3)$ and lattice symmetry breaking.
In this phase, the fermion spectrum is fully gapped.
On the level of mean-field theory, the coexistence state apparently gains energy by gapping out the remaining gapless fermion mode of the antiferromagnetic state through developing a weak bond anisotropy, in a way that it remains energetically favorable to break both $\SO(3)$ and lattice symmetry simultaneously, instead of breaking the lattice symmetry only.
For even $N$, the fermion spectrum is already fully gapped in the antiferromagnetic phase, such that the energetic competition drives a direct transition between the pure antiferromagnetic and dimerized orders, without a coexistence phase.
We stress that the occurrence of the coexistence phase for odd $N$ hinges on the precise energetic competition of the different microscopic states involved.
Indeed, from the above argument, one may have expected an analogous coexistent order to be realized for $N=5$ on the $\pi$-flux lattice, where for intermediate interaction, antiferromagnetic order with a remnant gapless Majorana band is realized.
However, our mean-field results on the $\pi$-flux lattice for $N=5$ instead suggest a direct transition between the antiferromagnetic and dimerized orders, without a coexistence phase, see Fig.~\ref{fig:pd}(b).

\section{Quantum phase transitions}
\label{sec:RG}

In this section, we discuss the natures of the quantum phase transitions occurring between the different phases.

\subsection{Semimetal-to-dimerized-phase transition}

On the mean-field level, the transition from the Majorana semimetal towards the dimerized phase, across which part of the lattice symmetry gets spontaneously broken, is strongly first order.
As this transition occurs at large values of $N$, for which quantum fluctuations around the mean-field solution are suppressed, we expect this conclusion to hold also beyond mean-field theory.

\subsection{Antiferromagnet-to-dimerized-phase transition}

For intermediate values of $N$, a transition from the $\SO(N)$-symmetry-breaking antiferromagnet to the lattice-symmetry-breaking dimerized phase becomes possible. These states break different symmetries, and a direct transition between them is therefore expected to be first order on general grounds, unless nontrivial fractionalized excitations play a role at the transition~\cite{senthil04}.
This expectation is consistent with our mean-field result for $N\geq 4$, in which case we indeed find a direct first-order transition.
For $N = 3$ on the honeycomb lattice, the mean-field calculations suggest a transition towards a coexistence phase, which in principle could be continuous, but turns out to be first order on the mean-field level. It should be emphasized, however, that the mean-field calculation may be uncontrolled at this small value of $N$. We leave the study of the effects of quantum fluctuations at this transition for future work.
For $N=2$ on both lattices, we find the generalized antiferromagnet to be stable up to the large $J/t$ limit, in agreement with previous works~\cite{wang14, li15, motruk15, capponi15, scherer15, hesselmann16, huffman17, huffman20, schuler21}. For this small value of $N$, there is therefore no transition towards dimerized order.

\subsection{Semimetal-to-antiferromagnet transition}

The transition between the Majorana semimetal and the $\SO(N)$-symmetry-breaking antiferromagnet is continuous on the mean-field level and occurs at small $N$. Fluctuations may therefore play a significant role at this transition, and we employ a renormalization group analysis to study their effects.

\subsubsection{Gross-Neveu-SO(N) field theory}

To this end, we reinterpret the $N(N-1)/2$ mean-field parameters $\phi^{ab}$, $1\leq a < b \leq N$, as off-diagonal components of a real antisymmetric tensor $(\phi^{ab}) \in \mathbb{R}^{N\times N}$, which henceforth is promoted to a fluctuating tensor-order-parameter field.
Using a gradient expansion, the microscopic model in the vicinity of the semimetal-to-antiferromagnet transition can be mapped onto a corresponding continuum field theory, described by the Euclidean action $S = \int \rmd^d x \rmd \tau \mathcal L$ with Lagrangian
\begin{align}
\mathcal L & = 
\bar\psi_\alpha \gamma_\mu \partial_\mu \psi_\alpha
+ \frac14 \phi^{ab}\left(r-\partial_\mu^2\right)\phi^{ab}
+ \frac{g}{2} \phi^{ab} \bar\psi_\alpha (L^{ab})_{\alpha\beta} \psi_\beta 
\nonumber\\&\quad
+ \frac{\lambda_1}{4}\left(\phi^{ab}\phi^{ab}\right)^2
+ \lambda_2 \phi^{ab}\phi^{bc}\phi^{cd}\phi^{da},
\label{eq:lagrangian}
\end{align}
where $\psi_\alpha$ and $\bar\psi_\alpha \equiv \psi^\dagger_\alpha\gamma_0$ with $\alpha = 1,\dots,N$ are $N$ colors of Dirac fermion fields that arise from combining each pair of lattice Majorana operators per color at the two valleys in the Brillouin zone of the microscopic model into a complex fermion field~\cite{seifert20a}. 
In the physical model with two sublattices in $d=2$ spatial dimensions, the fermion fields have $d_\gamma = 2$ components per color, where $d_\gamma$ corresponds to the dimension of the Clifford algebra representation, $\{\gamma_\mu,\gamma_\nu\} = 2\delta_{\mu\nu} \mymathbb{1}_{d_\gamma}$, $\mu,\nu \in \{0,\dots,d\}$.
However, for the interpretation of our results, it will prove convenient to compute the renormalization group flow for general $d_\gamma \in \mathbb{R}$ and postpone setting $d_\gamma = 2$ until later. This allows us to identify the effects of the fermion fluctuations on our results explicitly by smoothly interpolating from $d_\gamma = 0$ towards its physical value.
We will also consider general spatial dimension $1<d<3$, allowing us to perform an $\epsilon = 3-d$ expansion around the unique upper critical spatial dimension of three.
%
%
In Eq.~\eqref{eq:lagrangian}, we have assumed summation convention over all $a,b,c,d \in \{1,\dots,N\}$ using $L^{ba} = - L^{ab}$ and $L^{11} = \dots = L^{NN} = 0$, such that $\phi^{ab}\phi^{ab} \equiv - \Tr(\phi^2)$ and $\phi^{ab}\phi^{bc}\phi^{cd}\phi^{da} \equiv \Tr(\phi^4)$.

The order-parameter field can be understood to arise from a Hubbard-Stratonovich transformation of the four-fermion term $\left[\bar\psi_\alpha (L^{ab})_{\alpha\beta}\psi_\beta\right]^2$, and as such couples linearly to the corresponding fermion bilinear, with Yukawa coupling $g$.
Fermion fluctuations will generate a kinetic term for $\phi$, which has therefore been included from the outset in Eq.~\eqref{eq:lagrangian}.
The parameter $r$ acts as a tuning parameter for the transition: On tree level, we have $\langle \phi^{ab} \rangle = 0$ for $r>0$, corresponding to the Majorana semimetal phase, while $\langle \phi^{ab} \rangle \neq 0$ for $r<0$, corresponding to the $\SO(N)$-symmetry-broken antiferromagnetic phase.
A finite expectation value for $\phi^{ab}$ will at least partly gap out the Majorana spectrum.
From our mean-field results, we expect that the $\SO(N)$-symmetry-breaking ground state 
%
%
%
realizes the ``maximal'' symmetry breaking pattern of Eq.~\eqref{eq:pattern}, corresponding to a ``maximal'' gapping of the Majorana spectrum, and leaving behind at most only a single gapless mode.

Fluctuations will also generate bosonic self-interactions, which have therefore also been included in Eq.~\eqref{eq:lagrangian}. In the present situation with an antisymmetric real order-parameter field, there are in general two different interactions possible, parametrized by the couplings $\lambda_1$ and $\lambda_2$ in Eq.~\eqref{eq:lagrangian}. This is because $[\Tr(\phi^2)]^2$ and $\Tr(\phi^4)$ are generically independent for $N \geq 4$.
The fact that the space of bosonic self-interaction is spanned by two couplings $\lambda_1$ and $\lambda_2$ for $N\geq 4$ has important consequences for the fixed-point structure, as we shall see below.
The cases $N=2$ and $N=3$, however, are special: In these cases, we have $[\Tr(\phi^2)]^2 \equiv 2\Tr(\phi^4)$, such that the bosonic self-interaction terms can be written as
\begin{align} \label{eq:bosonic-interaction-N-2-3}
\frac{\lambda_1}{4}\left(\phi^{ab}\phi^{ab}\right)^2
+ \lambda_2 \phi^{ab}\phi^{bc}\phi^{cd}\phi^{da}
= 
\frac{\lambda_1 + 2 \lambda_2}{2}\left[\Tr(\phi^2)\right]^2,
\end{align}
for $N\leq 3$.
Within the naming scheme of Ref.~\cite{seifert20a}, the theory defined by Eq.~\eqref{eq:lagrangian} may be dubbed Gross-Neveu-SO($N$), as the fermion bilinear, to which the order parameter $\phi$ couples, transform in the fundamental representation of SO($N$).

\subsubsection{Renormalization group flow}

Integrating out the momentum and frequency modes in a thin shell between $\Lambda/b$ and $\Lambda$ in the $(d+1)$-dimensional reciprocal space, where $\Lambda$ is the ultraviolet cutoff, leads to the flow equations at one-loop order
\begin{align}
\frac{\rmd g^2}{\rmd \ln b} & = (\epsilon - \eta_\phi - 2\eta_\psi)g^2 - 2g^4, \label{eq:flow-g2} \\
\frac{\rmd \lambda_1}{\rmd \ln b} & = (\epsilon - 2\eta_\phi) \lambda_1 - 2(N^2 - N + 16) \lambda_1^2
\nonumber\\&\quad
- 16(2N-1)\lambda_1 \lambda_2 - 96\lambda_2^2, \label{eq:flow-lambda1} \\
\frac{\rmd \lambda_2}{\rmd \ln b} & = (\epsilon - 2\eta_\phi) \lambda_2 - 48 \lambda_1 \lambda_2 - 8(2N-1)\lambda_2^2
+ \frac{d_\gamma}{4}g^4, \label{eq:flow-lambda2} 
\end{align}
with the anomalous dimensions
\begin{align} \label{eq:eta}
\eta_\phi & = d_\gamma g^2, &
\eta_\psi & = \frac{N-1}{2} g^2,
\end{align}
and where we have rescaled the couplings as $(g^2,\lambda_1,\lambda_2) /(8\pi^2\Lambda^\epsilon) \mapsto (g^2,\lambda_1,\lambda_2)$, with $\epsilon = 3-d$ being the deviation from the upper critical dimension.
In order to arrive at the above flow equations, we have used a set of identities for the SO($N$) generators, which are discussed in Appendix~\ref{app:identities}.
These flow equations allow a number of crosschecks:
We have verified that for $g^2 = 0$, the flows of the bosonic self-interactions $\lambda_1$ and $\lambda_2$ are equivalent to those of the purely bosonic O($N$)-symmetric model with antisymmetric tensor order parameter~\cite{antonov13, antonov17}.
For $N=2$ and $N=3$, there is a only one quartic bosonic self-interaction term, parametrized by the coupling $\lambda \equiv \lambda_1 + 2\lambda_2$. Using Eqs.~\eqref{eq:flow-lambda1} and \eqref{eq:flow-lambda2}, we find in this case
\begin{align}
    \frac{\rmd \lambda}{\rmd \ln b} & = (\epsilon - 2\eta_\phi)\lambda
    - 2\left[N(N-1) + 16\right] \lambda^2
    + \frac{d_\gamma}{2} g^4.
\label{eq:flow-lambda-N-2-3}
\end{align}
For $N=2$, Eqs.~\eqref{eq:flow-g2}, \eqref{eq:eta}, and \eqref{eq:flow-lambda-N-2-3} are equivalent, modulo a rescaling of the couplings, to the flow equations of the Gross-Neveu-Ising model in $3-\epsilon$ spatial dimensions~\cite{zinnjustin91, karkkainen94, herbut09, janssen14, zerf17, janssen18}.
For $N=3$, they are in agreement with the previous explicit calculation for the Gross-Neveu-SO(3) model in Refs.~\cite{seifert20a, ray21}.%
\footnote{Note that in Refs.~\cite{seifert20a, ray21}, the variable $N$ denotes the number of two-component fermion flavors for fixed internal symmetry group SO(3), which corresponds to $3 d_\gamma/2$ in this work.}

\subsubsection{Continuous transition for \(\mathit{N\leq 3}\)}

For $N \leq 3$, there are only two independent couplings $g^2$ and $\lambda \equiv \lambda_1 + 2 \lambda_2$. Their flow equations \eqref{eq:flow-g2}, \eqref{eq:eta}, and \eqref{eq:flow-lambda-N-2-3} admit an infrared stable fixed point at
\begin{align}
    g^2_\star & = \frac{\epsilon}{N + d_\gamma + 1} + \mathcal O(\epsilon^2), \\
    \lambda_\star & = \frac{N+1 - d_\gamma + f(d_\gamma,N)}{4 (N + d_\gamma + 1) \left[N(N-1)+16\right]}
    \epsilon + \mathcal O(\epsilon^2),
\end{align}
where $f(d_\gamma,N) \equiv d_\gamma \sqrt{1+\frac{4 N^2-6N+62}{d_\gamma}+\left(\frac{N+1}{d_\gamma}\right)^2} > d_\gamma$, such that both $g_\star^2 >0$ and $\lambda_\star > 0$. The stable fixed point corresponds to a quantum critical point in the Gross-Neveu-SO($N$) universality class for $N \leq 3$.
The semimetal-to-antiferromagnet transition in the SO(2) and SO(3) Majorana-Hubbard models is therefore continuous, in agreement with the mean-field result of Sec.~\ref{sec:MFT}.
The quantum critical behavior is characterized by a set of universal critical exponents. The fermion and boson anomalous dimensions are readily obtained from Eqs.~\eqref{eq:eta} as
\begin{align}
    \eta_\phi & = \frac{d_\gamma \epsilon}{N+ d_\gamma + 1} + \mathcal O(\epsilon^2), &
    \eta_\psi & = \frac{(N-1)\epsilon}{2 (N + d_\gamma + 1)} + \mathcal O(\epsilon^2).
\end{align}
The correlation-length exponent $\nu$ is determined from the flow of the tuning parameter $r$,
\begin{align}
    \frac{\rmd r}{\rmd \ln b} = (2 - \eta_\phi) r + 2 \left[N(N-1) + 4\right] \frac{\lambda}{1+r} - 2 d_\gamma g^2,
\end{align}
yielding
\begin{align}
    1/\nu & = 2 - \biggl\{ \frac{d_\gamma}{N + d_\gamma+1}
    \nonumber \\ &\quad
    + \frac{[N + 1 -d_\gamma+f(d_\gamma, N)][N(N-1)+4]}{2 (N + 1 + d_\gamma)[N(N-1)+16]} \biggr\} \epsilon
    \nonumber \\ &\quad
    + \mathcal O(\epsilon^2).
\end{align}
As there appears no dangerously irrelevant coupling in the problem, we expect hyperscaling to hold. The exponents $\alpha$, $\beta$, $\gamma$, and $\delta$ can then be obtained from $\eta_\phi$ and $\nu$ by making use of the standard scaling relations~\cite{herbutbook}.
We note in passing that the above values for the critical exponents agree with those for the Gross-Neveu-Ising model~\cite{zinnjustin91, karkkainen94, herbut09, janssen14, zerf17, janssen18} upon setting $N=2$ and the Gross-Neveu-SO(3) model~\cite{seifert20a} for $N=3$, as expected.
In the latter case, recent results from three-loop and other elaborate approximations~\cite{ray21} show that while the values of the critical exponents receive quantitative corrections for finite $\epsilon$, the main conclusion concerning the nature of the transition holds also beyond the one-loop order.
The same is true for the $N=2$ case, for which results up to four-loop order are available~\cite{zerf17}.

\subsubsection{Fluctuation-induced first-order transition for \(\mathit{N\geq 4}\)}

For $N \geq 4$, the presence of two independent bosonic self-interaction terms is crucial. In order to identify the physical mechanism that leads to the absence of a stable fixed point in the parameter space spanned by $g^2$, $\lambda_1$, and $\lambda_2$ in the coupled fermion-boson field theory, it is useful to analytically continue the flow equations to noninteger dimension $d_\gamma$ of the Clifford algebra representation, $0 \leq d_\gamma \leq 2$, where the limit $d_\gamma = 2$ corresponds to the physical situation in $d=2$ spatial dimensions. This allows us to smoothly incorporate the effects of fermion fluctuations.
We start by discussing the $N=4$ case explicitly. The generalization to $N>4$ will be discussed afterwards.

\begin{figure*}[tp]
\centering
\includegraphics[width=\textwidth]{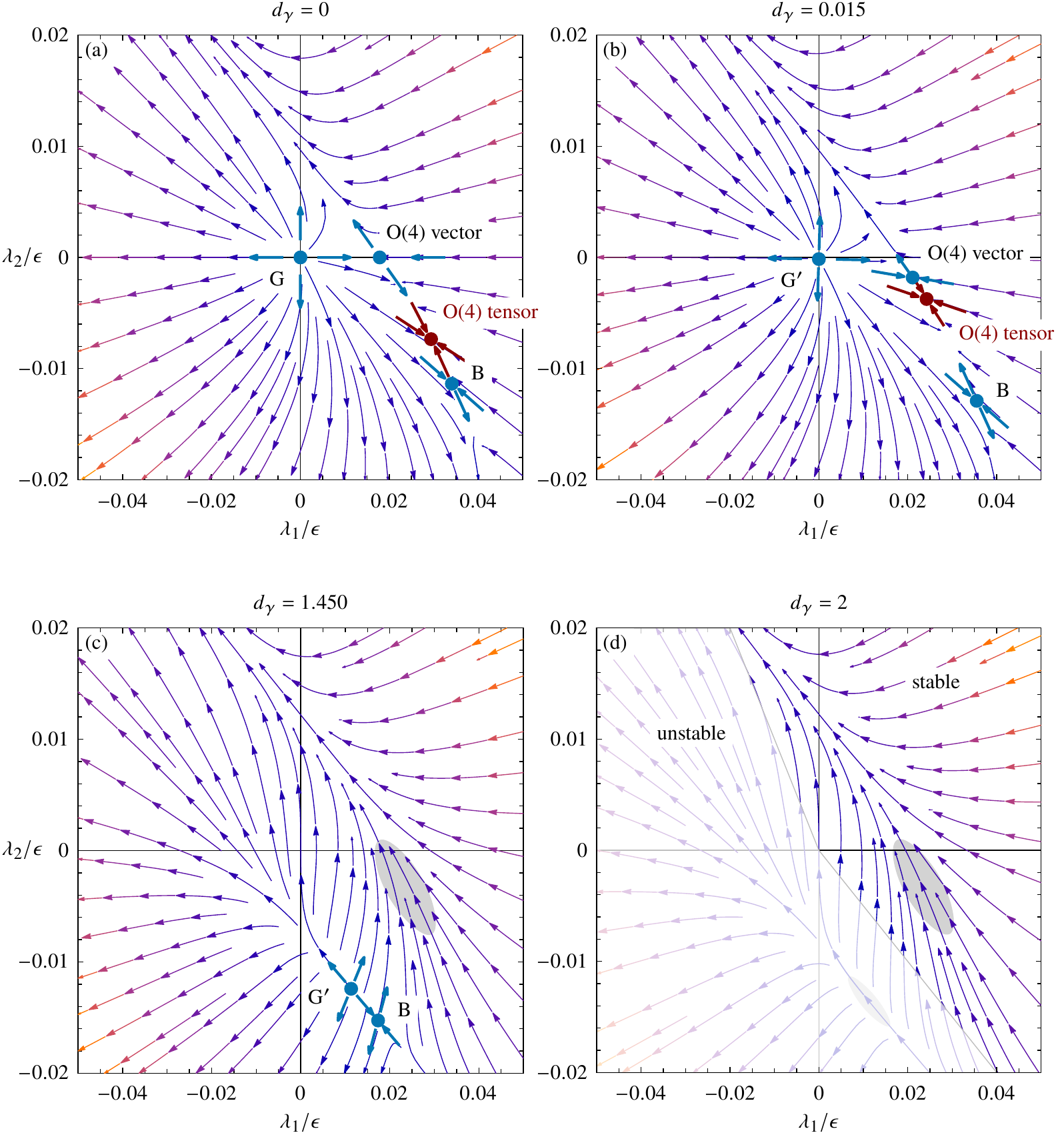}
\caption{Evolution of fixed-point structure of Gross-Neveu-SO(4) field theory as function of dimension of Clifford algebra representation $d_\gamma$, analytically continued to noninteger values. Arrows denote flow towards infrared.
(a) Bosonic case $d_\gamma = 0$, featuring, besides the Gaussian fixed point G, three interacting fixed points: The O(4) vector fixed point at $\lambda_2 = 0$ and the fixed point B are bicritical, while the O(4) tensor fixed point at finite $\lambda_1 > 0$ and $\lambda_2 < 0$ is fully infrared stable, corresponding to a continuous phase transition.
(b) Upon the successive inclusion of fermion fluctuations, the O(4) vector and O(4) tensor fixed points approach each other as function of increasing $0 < d_\gamma < 0.0164$. The counterpart of the Gaussian fixed point G$'$ has moved slightly towards finite interaction. Here, $d_\gamma = 0.0150$, i.e., slightly below $d_\gamma^\text{cr,1}$.
(c) For $0.0165 < d_\gamma < 1.4853$, the O(4) vector and O(4) tensor fixed points have annihilated, leaving behind a regime of slow flow in parameter space (gray shaded region). Here, $d_\gamma = 1.450$, i.e., slightly below $d_\gamma^\text{cr,2}$.
(d) For $d_\gamma > 1.4854$, fixed points B and G$'$ have annihilated as well, leaving behind only the runaway flow towards $\lambda_1 \to -\infty$ and $\lambda_2 < |\lambda_1|$ (faint region marked as ``unstable''), suggesting a weak first-order transition. Here, $d_\gamma = 2$, corresponding to the physical situation in $d=2$ spatial dimensions.}
\label{fig:rgflow}
\end{figure*}

In the limit $d_\gamma = 0$, the Yukawa coupling $g$ and the pair of bosonic self-couplings $(\lambda_1, \lambda_2)$ completely decouple, such that the flow in the bosonic sector is equivalent to those of purely bosonic model studied in Refs.~\cite{antonov13, antonov17}. The fixed-point structure of this model is depicted for $N=4$ in Fig.~\ref{fig:rgflow}(a).
Besides the Gaussian fixed point at $(\lambda_1,\lambda_2) = (0,0)$, there are three interacting fixed points: One of it is located at $\lambda_1 > 0$ and $\lambda_2 = 0$, and it corresponds to the usual Wilson-Fisher fixed point in the O(4) vector model. We denote it as ``O(4) vector'' in Fig.~\ref{fig:rgflow}(a). Importantly, in the  vicinity of this fixed point, $\lambda_2$ corresponds to an infrared relevant parameter, rendering the O(4) vector fixed point bicritical.
Another bicritical fixed point is located at $\lambda_1 > 0$ and $\lambda_2 < 0$, denoted as ``B'' in Fig.~\ref{fig:rgflow}(a).
There is, however, a unique critical fixed point that is fully infrared stable in the $\lambda_1$-$\lambda_2$ coupling plane, and thus corresponds to a continuous transition in the purely bosonic model. As the order parameter $\phi^{ab}$ is a real antisymmetric tensor, we denote this fixed point as ``O(4) tensor.''

Upon the inclusion of fermion fluctuations in the flow of $(\lambda_1, \lambda_2)$ for finite, but small, $d_\gamma > 0$, all four fixed points move within the coupling plane. In particular, the O(4) vector and O(4) tensor fixed points approach each other, Fig.~\ref{fig:rgflow}(b), and collide at $d_\gamma^\text{cr,1} = 0.0164$. For $d_\gamma > d_\gamma^\text{cr,1}$, the O(4) tensor and O(4) vector fixed points have annihilated and moved into the complex coupling plane. A similar fixed-point-annihilation scenario has previously been discussed in a number of relativistic~\cite{halperin74, nahum15, ihrig19, gies06, kaplan09, braun14, herbut16b, herbut16a, gracey18, ma19, nahum20, ma20} and nonrelativistic~\cite{herbut14, janssen17} field theories.
For values of $d_\gamma$ above, but not too far from $d_\gamma^\text{cr,1}$, the imaginary parts of the complex fixed-point values are small. This implies that the renormalization group flow in the real coupling plane slows down in the regime where the two fixed points have annihilated. This regime is indicated as gray shaded region in Fig.~\ref{fig:rgflow}(c). The slow flow induces an exponentially large, but finite correlation length $\xi \propto \rme^{A/\sqrt{d_\gamma - d_\gamma^\text{cr,1}}}$, where $A>0$ is a dimensionless constant of order unity~\cite{kaplan09, braun11, herbut14}. 
The situation can be understood as an example of a fluctuation-induced first-order transition, in analogy to the case of the Abelian Higgs model~\cite{halperin74, nahum15, ihrig19}.
Upon further increasing $d_\gamma$, we find a second critical $d_\gamma^\text{cr, 2} = 1.4853$, at which the two remaining fixed points collide, and annihilate as well for $d_\gamma > d_\gamma^\text{cr,2}$.

For the case of $d_\gamma = 2$, corresponding to the physical situtation in two spatial dimensions, there are no fixed points left, leaving behind only the runaway flow towards $\lambda_1 \to - \infty$ and $\lambda_2 < |\lambda_1|$, see Fig.~\ref{fig:rgflow}(d).
Local stability of the effective order-parameter potential $V(\phi^{ab})$ near criticality $r = 0$ and small $\phi^{ab}$ requires
\begin{align} \label{eq:stability}
    V(\phi^{ab})\bigr|_{r = 0} = \frac{\lambda_1}{4} \left(\phi^{ab} \phi^{ab} \right)^2 + \lambda_2 \phi^{ab}\phi^{bc}\phi^{cd}\phi^{da} > 0
\end{align}
for all real antisymmetric matrices $(\phi^{ab}) \in \mathbb R^{4\times 4}$.
By making use of the O(4) symmetry, we can rotate any field configuration into a block-diagonal form
\begin{align}
    \phi \mapsto O^\top \phi O = 
    \begin{pmatrix}
    0 & m_1 & 0 & 0 \\
    -m_1 & 0 & 0 & 0 \\
    0 & 0 & 0 & m_2 \\
    0 & 0 & -m_2 & 0
    \end{pmatrix}
\end{align}
with orthogonal matrix $O \in \mathbb R^{4 \times 4}$ and real $m_1, m_2 \in \mathbb R$. It is then straightforward to show that the local-stability criterion~\eqref{eq:stability} is equivalent to
\begin{align}
    \lambda_2 >
    \begin{cases}
    |\lambda_1| & \text{for } \lambda_1 < 0,\\
    - |\lambda_1|/2 & \text{for } \lambda_1 > 0.\\
    \end{cases}
\end{align}
The faint region in Fig.~\ref{fig:rgflow}(d) indicates the values of $(\lambda_1,\lambda_2)$ for which the above criterion is violated, marked as ``unstable.'' In particular, Fig.~\ref{fig:rgflow}(d) illustrates that the quartic boson self-interaction couplings always flow towards the locally-unstable region in the infrared, independent of their ultraviolet starting values. Higher-order terms beyond the quartic order are therefore required to ensure global stability of the effective potential, and for $r$ above, but close to zero, there must be a global minimum at finite $\phi$ that is lower in energy than the local minimum at $\phi = 0$.
Together with the exponentially large correlation length, this implies that the semimetal-to-antiferromagnet transition in the SO(4) Majorana-Hubbard model is discontinuous, but only very weakly so.

Let us now discuss the fixed-point structure for $N>4$. Again, we start with the bosonic case $d_\gamma = 0$ first and discuss the effects of fermion fluctuations for $d_\gamma > 0$ afterwards.
Furthermore, we now analytically continue the flow equations also for noninteger values of $N$, allowing us to track the evolution of the fixed points as function of this parameter as well, starting from the $N=4$ flow diagram in Fig.~\ref{fig:rgflow}(a).
Interestingly, we now find already for fixed $d_\gamma = 0$ as function of increasing $N > 4$ a fixed-point-annihilation scenario that is very similar to the one discussed above for fixed $N=4$ as function of increasing $d_\gamma>0$.
The main difference to the situation for fixed $N=4$ is that the stable O(4) tensor fixed point now collides and annihilates with the bicritical fixed point B as function of $N>4$ for fixed $d_\gamma = 0$. At the one-loop order, this happens already at $N^\text{cr}(d_\gamma = 0) = \frac{2 + 3\sqrt{22}}{4} = 4.0178$.
For $N > N^\text{cr}$, no infrared stable fixed point remains for real values of the couplings, leaving behind the runaway flow now already for $d_\gamma = 0$.
Fermion fluctuations for finite $d_\gamma$ do not bring these two fixed points back into the real coupling plane, as we have explicitly verified by numerically solving the fixed-point equations for various values of $N$ and $d_\gamma$. Instead, increasing $d_\gamma > 0$ for fixed $N > N^\text{cr}(d_\gamma=0)$ leads to a collision of the remaining two unstable fixed points, similar to the situation for $N=4$ shown in Figs.~\ref{fig:rgflow}(b)--(d).
As a consequence, the semimetal-to-antiferromagnet transition in the SO($N$) Majorana-Hubbard model is discontinuous for all values of $N \geq 4$, for which the antiferromagnetic phase can be stabilized on a given lattice.
As the flow is the slower the smaller the value of $N$, the fluctuation-induced first-order transition will be weakest for $N=4$, and less weak for larger~$N$.

Let us comment on what might be expected from corrections beyond the one-loop approximation.
The bosonic sector for $d_\gamma = 0$ has recently been discussed at the three-loop order in Ref.~\cite{antonov17}. These results suggest that the two fixed points at finite values of $\lambda_2$ in Fig.~\ref{fig:rgflow}(a) exchange stability as function of $\epsilon$ above a certain finite critical value, which is estimated to be below the physical value of one.
This would imply that in $d=2$ spatial dimensions, the infrared stable fixed point is the one denoted as ``B'' in Fig.~\ref{fig:rgflow}(a).
The fact that this fixed point annihilates, in the one-loop approximation, at a significantly larger critical value $d_\gamma^\text{cr,2} \gg d_\gamma^\text{cr,1}$, still leaves room for the existence of a stable fixed point at $\epsilon = 1$, $d_\gamma = 2$, and $N=4$.
In this scenario, the $N=4$ transition would then still be continuous.
For larger values of $N$, however, the imaginary parts of the complex fixed-point couplings are sizable at the one-loop order, inhibiting the revival of these fixed points in the real coupling plane upon the inclusion of higher-loop corrections.
We therefore believe that our result of a fluctuation-induced first-order transition that occurs beyond a certain critical value of $N$ is a robust feature of the SO($N$) Majorana-Hubbard models in $d=2$ spatial dimensions, although the value of $N$ above which this happens might receive corrections beyond the one-loop order.
Estimating these corrections represents an interesting direction for future work.

\section{Conclusions}
\label{sec:conclusions}
%
In conclusion, we have studied SO($N$) Majorana-Hubbard models on honeycomb and $\pi$-flux square lattices in the zero-temperature limit.
On both lattices, the phase diagrams feature three symmetry-distinct phases: For weak interactions, the disordered Majorana semimetal is stable for all values of $N$. For strong interactions above a certain finite threshold, however, an ordered state that breaks SO($N$) symmetry and can be understood as a N\'eel antiferromagnet is stabilized when $N$ is small, while a dimerized state that breaks the lattice symmetry and can be understood as a staggered valence bond solid is found for large $N$.
These results are reminiscent of the situation of complex fermions in the SU($N$) Hubbard-Heisenberg model~\cite{affleck88, marston89, read89, read90, assaad05, lang13}: There, in the large-$N$ limit and for half filling, a Dirac semimetal phase is stabilized for weak interactions (dubbed ``flux'' phase in Refs.~\cite{affleck88, marston89}), which gives way to a dimerized ``spin-Peierls'' phase (also dubbed ``columnar valence bond solid'' in Ref.~\cite{lang13}) upon increasing the interaction strength above a finite threshold. 
For small $N$, on the other hand, the N\'eel antiferromagnet occurs in the limit of strong interactions~\cite{read89, read90}. Further exotic symmetry-broken states may be stabilized for intermediate values of $N$ and strong interactions as well~\cite{assaad05, lang13, beach09, toth10, corboz11}.
The phase diagrams of our SO($N$) Majorana-Hubbard models feature an analogous structure, with the Majorana semimetal replacing the Dirac semimetal at weak interaction, the staggered dimerized phase replacing the columnar dimerized phase at strong interaction and large $N$, and the N\'eel antiferromagnet occurring in both families of models at strong interaction and small $N$.

In this work, we have restricted ourselves to translation-invariant states. Note that this ansatz excludes states with enlarged unit cells, such as columnar or plaquette valence bond solids. From the analogy between $\SO(N)$ Majorana-Hubbard and $\SU(N)$ Hubbard-Heisenberg models, it seems possible that the staggered dimerized state could be replaced, in some parts of the phase diagram, by other states that feature larger unit cells, if the latter are taken into account in a refined mean-field analysis. We leave this interesting question for future work.

\begin{figure*}
    \centering
    \includegraphics[width=.85\textwidth]{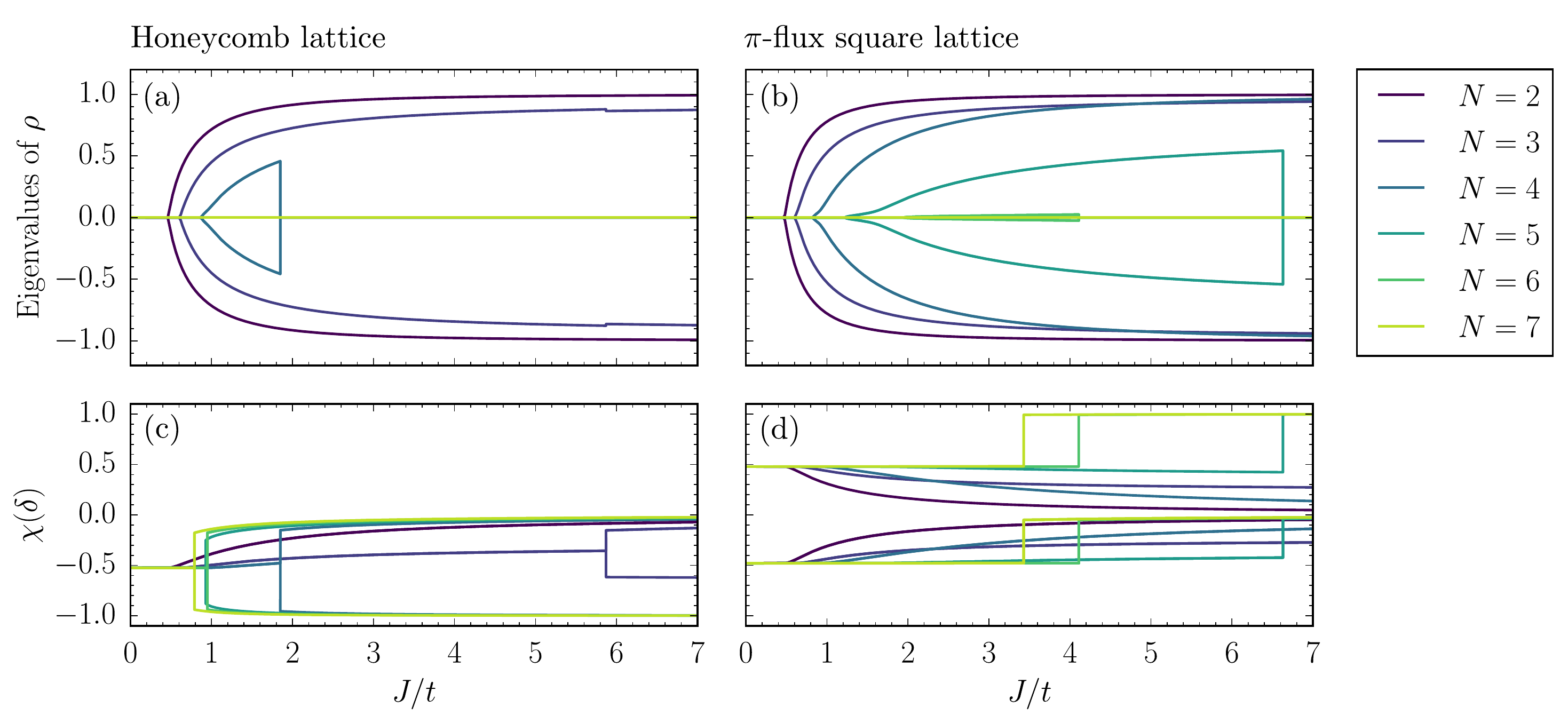}
    \caption{Mean-field parameters of SO($N$) Majorana-Hubbard model as function of $J/t$ for different values of $N$ on (a,c) honeycomb and (b,d) $\pi$-flux square lattices. (a,b) Eigenvalues $\pm \tilde \phi$ of antiferromagnetic order parameter matrix $\rho = \sum_{a<b} \phi^{ab} L^{ab}$. (c,d) Dimer order parameters $\chi(\delta) \in \{\tilde \chi - \Delta \chi, \tilde \chi\}$ on different nearest-neighbor bonds $\delta$ in $\langle ij\rangle = \langle i, i+\delta\rangle$.
    }
    \label{fig:mftparameters}
\end{figure*}

For the SO($N$) Majorana-Hubbard models, we have shown that the transition between the symmetric Majorana semimetal and the SO($N$)-symmetry-broken N\'eel antiferromagnet is continuous for small $N$, but becomes discontinuous for $N$ above a certain critical value. In the one-loop approximation, the fluctuation-induced first-order transition occurs for $N\geq4$, while the cases for $N \leq 3$ feature a continuous transition.
Higher-loop corrections may shift the critical value of $N$, above which the transition turns first order; however, the qualitative features of the one-loop result are expected to hold also beyond our approximation.
Recently, a lattice model, which is amenable to sign-problem-free quantum Monte Carlo simulations, and is expected to feature a quantum critical point in the universality class of the semimetal-to-SO(3)-antiferromagnet transition, has been devised~\cite{liu21}. The numerical data obtained for this model are consistent with a continuous transition, in agreement with our results for the SO(3) Majorana-Hubbard model.
It would be interesting to generalize this model to larger values of $N$ in order to test our prediction of a fluctuation-induced first-order transition by means of large-scale numerical simulations.

\begin{acknowledgments}
%
We thank H.-H. Tu for illuminating discussions.
This work has been supported by the Deutsche Forschungsgemeinschaft (DFG) through SFB 1143 (A07, Project No.~247310070), the W\"{u}rzburg-Dresden Cluster of Excellence {\it ct.qmat} (EXC 2147, Project No.~390858490), and the Emmy Noether program (JA2306/4-1, Project No.~411750675), as well as in part by the National Science Foundation (NSF) under Grant No.~NSF PHY-1748958.
\end{acknowledgments}

\appendix
\setcounter{equation}{0}  
\renewcommand\theequation{A\arabic{equation}}

\section{Explicit forms of mean-field parameters}\label{app:mft}

In the main text, only the magnitudes of the antiferromagnetic and dimer order parameters are shown.
In this appendix, we discuss the various mean-field parameters explicitly as function of $J/t$, see Fig.~\ref{fig:mftparameters}.
The top panels of this figure show the eigenvalues of the antiferromagnetic order parameter $\phi^{ab} L^{ab}$ for various values of $N$, illustrating the continuous nature of the semimetal-to-antiferromagnet transition at the mean-field level, the discontinuous nature of the antiferromagnet-to-dimerized transition, and the absence of antiferromagnetic order at large $N \geq 5$ ($N \geq 7$) on the honeycomb ($\pi$-flux) lattice.
The bottom panels show the bond variables $\chi_{ij}$ on different nearest-neighbor bonds $\langle ij\rangle$ for various values of $N$. 
In the Majorana semimetal phase, all bond variables acquire the same $N$- and $J/t$-independent values $\chi_{ij} = -0.52483184$ and $\chi_{ij} = \mp 0.47902212$ on the honeycomb and $\pi$-flux lattices, respectively.
In the antiferromagnetic phase, the bond variables become $N$- and $J/t$-dependent, but continue to have the same magnitude on the different bonds.
Eventually, in the dimerized phase, the bond variables acquire different values on the different nearest-neighbor bonds. This leads to a splitting of the curves for $\chi_{ij}$ above a certain value of $J/t$, which indicates the antiferromagnet-to-dimerized and semimetal-to-dimerized transitions for intermediate $N$ and large $N$, respectively. Both transitions are characterized by sizable jumps in the dimer order parameter, indicating strong first-order transitions.

\section{Identities for \(\boldsymbol{\SO(N)}\) generators}\label{app:identities}

In this appendix, we list some identities for the $\SO(N)$ generators $L^{ab}$, which we have used to derive the flow equations~\eqref{eq:flow-g2}--\eqref{eq:eta}.
An explicit representation is given by $L^{ab}_{\alpha \beta} = -\rmi (\delta_\alpha^a \delta_\beta^b - \delta^a_\beta \delta^b_\alpha )$, with Greek indices $\alpha, \beta = 1, \dots, N$.
In order to avoid double counting, we restrict the indices of $L^{ab}$ as $1 \leq a < b \leq N$.
For given $a<b$ and $c<d$, the $N\times N$ matrices $L^{ab}$ and $L^{cd}$ satisfy the defining $\SO(N)$ algebra
\begin{equation}
    \left[L^{ab},L^{cd} \right] = \left(-\rmi\right) \left(L^{ad} \delta^{bc} - L^{ac} \delta^{bd} +L^{bc} \delta^{ac} - L^{bd} \delta^{ac}  \right),
\end{equation}
where matrix multiplication is understood when Greek indices are suppressed.
Note that the explicit factor of $\rmi$ in the above relation is due to our choice of imaginary $L^{ab}$.

With the representation chosen above, the $\SO(N)$ Casimir in the fundamental representation reads
\begin{equation} \label{eq:SONcompleteness}
    \sum_{1 \leq a < b \leq N} L^{ab}_{\alpha \beta} L^{ab}_{\gamma \rho} = - \delta_{\alpha\gamma} \delta_{\beta \rho}+\delta_{\alpha \rho} \delta_{\beta \gamma}.
\end{equation}
The above identity can be understood as a completeness relation in the space of purely imaginary antisymmetric $N\times N$ matrices, in which the $L^{ab}$ form a basis.
Contracting $\beta$ with $\gamma$ in Eq.~\eqref{eq:SONcompleteness}, one obtains
\begin{align}
    \sum_{a < b} L^{ab} L^{ab} = (N-1) \mymathbb{1}_N,
\end{align}
and
\begin{align}
    \sum_{a < b} \Tr L^{ab} L^{ab} = N(N-1).
\end{align}
For the triangle diagram contributing to the renormalization of the Yukawa coupling, we further make use of the identity
\begin{equation}
    \sum_{c<d} L^{cd} L^{ab} L^{cd} = L^{ab},
\end{equation}
which similarly follows from Eq.~\eqref{eq:SONcompleteness}.

\bibliographystyle{longapsrev4-2}
\bibliography{Majorana-SON}

\end{document}